\documentclass[aps,twocolumn,showpacs]{revtex4}

\usepackage{amsmath}
\usepackage{graphicx}

\begin{document}

\preprint{}

\title[]{Laser-driven atom moving in a multimode cavity: strong
  enhancement of cavity-cooling efficiency}

\author{Peter Domokos}
\altaffiliation{On leave from Research Institute for Solid State Physics and
   Optics, Budapest, Hungary}
\author{Thomas Salzburger}%
\author{Helmut Ritsch}%
\affiliation{%
Institut for Theoretical Physics, Universit\"at Innsbruck,
Technikerstr.~25, A-6020 Innsbruck,\\ Austria}%

\begin{abstract}
  
  Cavity-mediated cooling of the center--of--mass motion of a
  transversally, coherently pumped atom along the axis of a high--Q
  cavity is studied.  The internal dynamics of the atomic dipole
  strongly coupled to the cavity field is treated by a
  non-perturbative quantum mechanical model, while the effect of the
  cavity on the external motion is described classically in terms of
  the analytically obtained linear friction and diffusion
  coefficients.  Efficient cavity-induced damping is found which leads
  to steady-state temperatures well-below the Doppler limit. We reveal
  a mathematical symmetry between the results here and for a similar
  system where, instead of the atom, the cavity field is pumped.  The
  cooling process is strongly enhanced in a degenerate multimode
  cavity. Both the temperature and the number of scattered photons
  during the characteristic cooling time exhibits a significant
  reduction with increasing number of modes involved in the dynamics.
  The residual number of spontaneous emissions in a cooling time for
  large mode degeneracy can reach and even drop below the limit of a
  single photon.
\end{abstract}


\maketitle

\section{Introduction}

The mechanical effect of light on atoms in high-Q cavities is currently being
the subject of intensive theoretical and experimental research. The large
interest that this field attracts is due to the complexity of the
strongly-coupled dynamics of a moving atom and a few-photon field, as compared
to the relatively well-understood atomic motion in external laser fields.
Various effective cooling and trapping schemes have recently been found, which
can drive atoms to low, sub-Doppler temperatures on a fast time scale. The use
of dynamical fields in optical cavities is also a promising approach to extend
the power of ordinary laser cooling methods to a large variety of atomic
species.  There are cooling mechanisms independent of the internal electronic
structure of the particle, which could enable us to achieve the principal goal
of cooling molecules.

In the simplest cavity-cooling schemes
\cite{mossberg,lewenstein,cirac,vuletic99,vuletic01} the cavity is
used as a passive element to taylor the atomic spontaneous emission
rate by altering the spectral mode density. For example, in the
pioneering work by Mossberg {\it et al.}  \cite{mossberg}, an atom
moving in a resonant laser standing-wave field can undergo a
Sisyphus-type cooling in coloured vacuum, e.g.\ the one created inside
a cavity. The spontaneous emission is then favoured, by selection in
frequency space, from the top of a potential hill into the bottom of a
potential well.  Similarly, Vuleti\'c {\it et al.\ }
\cite{vuletic01} pointed out that a two-photon
Doppler-effect can occur provided the atom tends to scatter
inelastically an incoming laser photon into a cavity mode of higher
resonance frequency.

Dynamic cavity-cooling effects have been first predicted in
Refs.~\cite{horak97,hechenblaikner98}. In these schemes the free atom
moves in a weakly driven cavity field. The damping relies on a
strongly coupled atom-field system, in which the field exhibits
non-adiabatic dynamics due to the finite mode bandwidth $\kappa$.  Let
us emphasize that the cooling mechanism is quite sophisticated,
requiring at least two elements: (i) strong atom-field coupling
manifested by the dressed state spectrum of the atom-cavity system;
(ii) a cavity bandwidth in a specific narrow parameter range for the
non-conservative dynamics.  Both conditions are present in current
setups \cite{pinkse00,hood00} and indications of the predicted
mechanical effects have been observed in experiments.

In this paper we will consider nonadiabatic effects in a configuration
similar to the ones in Refs.\ 
\cite{mossberg,lewenstein,vuletic99,vuletic01}, that is, a moving atom
is driven by a classical laser field from the side, and cavity photons
are created only via the atom, by scattering from the laser field.  We
proceed in two respects. First, we move towards the strong coupling
regime where the non-conservative dynamical effects, analogous to the
ones explored in Refs.\ \cite{horak97,hechenblaikner98}, appear and
amount to new cooling regimes in the parameter space.  Accordingly, we
will present a non-perturbative quantum treatment of the internal
dynamics of the coupled atom-cavity system. Our model is exact in the
low-saturation regime that happens either for large atomic detuning or
for weak external driving intensity. The diffusion coefficient due to
dipole heating and the linear friction coefficient will be
analytically calculated, which allows for the statistical description
of the center-of-mass (CM) motion characteristics. We obtain the
somewhat surprising result that pumping the atomic or pumping the
field component of the coupled system are not the same even for low
saturation. Although the similarity appears in the form of a quite
intuitive symmetry between the results we derive and the ones
presented in Refs.\ \cite{horak97,hechenblaikner98} for the
cavity-driven system, important physical differences follow from the
two different situations. Such consequences can be found, for example,
in the temperature dependence on the system parameters.

As a second step forward, we will study multimode effects of the
cavity-induced forces. Multimode fields have been conjectured to give
rise to a significant improvement in terms of steady-state temperature
when many cavity modes are driven simultaneously \cite{horak01}, and
also in the case of the two-photon Doppler effect \cite{vuletic01}. In
the physical configuration we consider in this paper, the only
component of the system being externally driven is the atom. It is a
very natural and, in practice, a simple extension to put a multimode
cavity, such as a confocal resonator, around the driven particle. We
ask what happens if there are several modes interacting with the
atomic dipole. Here we will concentrate on scaling laws rather than on
the geometrical aspects of the multimode cavity field, this latter is
being important, e.\ g., to reveal atomic trajectories
\cite{kaleidoscope}. We consider degenerate modes with closely uniform
mode functions. This situation proves to be an easy way to effectively
enhance the dipole interaction strength and reach previously
unexplored regimes of the dynamics. One of the main questions is if the
number of spontaneously scattered photons during the cooling time
can be restricted below the single photon level, which is the
necessary requirement for effectively cooling particles without closed
two-level pumping cycle.

The paper is organized as follows. In Sect.\ 2 we present the model of
the system, and introduce the definition of the mechanical forces and
the diffusion that we use to describe the semiclassical CM motion. In
Sect.\ 3 the set of linearized Heisenberg--Langevin equations is
solved which leads to the analytical expressions for the linear
friction and the diffusion coefficients.  These results are analyzed
first for a single-mode cavity in Sect.\ 4, with the aim of
identifying the cooling mechanisms relying on a passive cavity and
those originating from the strongly-coupled atom-cavity dynamics.
Then, in Sect.\ 5., we study how the physically relevant quantities,
such as the steady-state temperature, the cooling time, and the number
of spontaneously scattered photons scale with the number of degenerate
modes of a multimode cavity. Finally, we conclude in Sect.\ 6.

\section{The model}

\begin{figure}[htbp]
  \begin{center}
    \includegraphics[width=7cm]{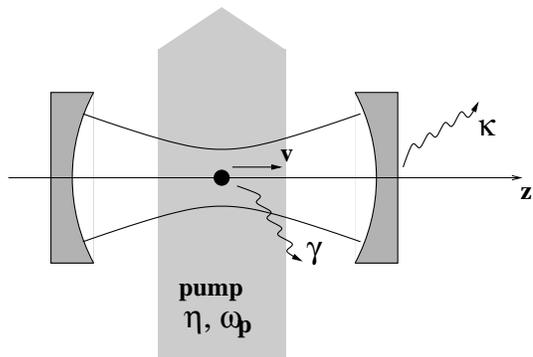}
\caption{Schematic representation of the system composed of a cavtity
  and a moving atom that is driven from the side.}
    \label{fig:scheme}
  \end{center}
\end{figure}
The system is composed of $M$ degenerate modes of a high-finesse
cavity interacting with a moving atom. The atom is driven from the
side by a coherent laser field (see Fig.\ \ref{fig:scheme}). We
restrict the study of the atomic motion to the longitudinal dimension
along the cavity axis. The driving field can be assumed uniform in the
relevant range of the atomic motion. This is equivalent to neglecting
the slow variation of the Gaussian envelope which is broad compared to
the wavelength.  The pumping strength is thus given by a single
parameter $\eta$.  The pumping frequency $\omega_p$ is detuned from
the atomic transition frequency with $\Delta_A=\omega_p-\omega_A$ and
from the common resonance frequency of the modes with
$\Delta_C=\omega_p-\omega_C$.  The dipole-cavity field interaction is
described by the Hamiltonian
\begin{multline}
  \label{eq:H}
  H = -\hbar \Delta_C \sum_{i=1}^M a_i^\dagger a_i - 
     \hbar \Delta_A \sigma^\dagger \sigma -i \hbar \eta
     (\sigma-\sigma^\dagger) \\ - i \hbar \sum_{i=1}^M 
     [g_i(z) \sigma^\dagger a_i - g_i^*(z) a_i^\dagger \sigma] \; ,
\end{multline}
which is written in a frame rotating at the pumping frequency
$\omega_p$. The atomic dipole operator is denoted by $\sigma$, and
$a_i$ is the bosonic field-amplitude operator associated with the
$i$th mode. The position-dependent dipole coupling is
$g_i=\sqrt{\hbar\omega_C/2\epsilon{\cal V}_i}\, d\, f_i(z)$, where $d$
is the atomic dipole moment, ${\cal V}_i$ is the effective volume of
the mode $i$, and $f_i(z)$ is the respective mode function which is
normalized to have the maximum equal to 1.

In addition to this coherent dynamics, the system is subject to
dissipation processes, i.e.\ to spontaneous emission with rate
$2\gamma$ into radiation modes other than the cavity ones, and to
cavity photon loss with a uniform rate of $2\kappa$ for all the
degenerate modes. The resulting open system can be described by
Heisenberg--Langevin equations.  We consider the low-saturation regime
where the atomic population inversion operator
$\sigma_z=(\sigma^\dagger\sigma-\sigma\sigma^\dagger)/2$ can be replaced
by the c-number $-1/2$. With this approximation one gets a linear set
of equations
\begin{subequations}
 \label{eq:HL}
\begin{align}
 \dot\sigma &= (i \Delta_A - \gamma) \sigma - \sum_{i=1}^M g_i(z) a_i
 +\eta + \xi_A \\
 \dot a_i &= (i\Delta_C - \kappa) a_i + g_i^*(z) \sigma + \xi_i \; ,
\end{align}
\end{subequations}
for all the modes $i=1..M$.  The Langevin noise operators are defined
by the second-order correlations
\begin{subequations}
\label{eq:L}
\begin{align}
\langle \xi(t_1) \xi^\dagger(t_2) \rangle &= 2 \gamma\,  \delta(t_1-t_2)\\
\langle \xi_i(t_1) \xi_j^\dagger(t_2) \rangle &= 2 \kappa\,  \delta_{ij}\, 
\delta(t_1-t_2)\; ,
\end{align}
\end{subequations}
all the other correlations vanish. Note that a simple transformation,
$\sigma \rightarrow \sigma -\eta/(i\Delta_A-\gamma)$ shifts the
driving from the atomic dipole $\sigma$ to the mode amplitudes $a_i$,
thus one could think that driving the field and driving the atomic
component leads to equivalent dynamics.  However, this is not true for
a moving atom when $z$ is considered also as a variable.

In the Heisenberg--Langevin equations of the field amplitudes and the
atomic dipole, the atomic position appears as a parameter.
Nevertheless, under the mechanical effect of the radiation field on
the atomic center--of--mass motion, the atomic position is also a
variable. In the present paper we use a consistent semiclassical model
that relies on a classical treatment of the CM motion \cite{cohen92}.
The external atomic CM dynamics is assumed to decouple from the
internal one given by the Eqs.~(\ref{eq:HL}), and it is governed by a
Langevin-type equation
\begin{subequations}
  \label{eq:CM_motion}
\begin{align}
  \dot x &= p/m \\
  \dot p &= f_p + \beta p + \xi_p \; ,
\end{align}
\end{subequations}
where $f_p$ is a conservative force, $\beta$ is a friction coefficient
and $\xi_p$ is a classical stochastic term giving rise to diffusion.
Statistical properties of the CM motion can be derived from this
Langevin-equation. This semiclassical approach is consistent as long
as the temperature is far above the recoil temperature $T_{\mbox{\tiny
    rec}}=\hbar^2 k^2/2m$, where $k$ is the wave number of the modes
involved, $m$ is the mass of the atom.

The aim of the present paper consists in calculating the coefficients
in Eq.~(\ref{eq:CM_motion}b). They all can be derived from the force
operator, which is given by the gradient of the interaction term in
the Hamiltonian \cite{cohen92}
\begin{equation}
  \label{eq:force}
  F = i\hbar \sum_{i=1}^M 
    [(\partial_z g_i) \sigma^\dagger a_i - (\partial_z g_i^*) a_i^\dagger \sigma]\; .
\end{equation}
First, the mean of the force operator provides directly $f_p$,
\begin{equation}
  \label{eq:mean_force}
  f_p = \langle F \rangle\; .
\end{equation}
Then, the noise term $\xi_p$ originates from the fluctuations of the
force operator. On evaluating the two-time correlation function,
\begin{equation}
  \label{eq:diffusion_def}
  \langle F(t_1) F(t_2) \rangle - \langle F(t_1)\rangle\langle F(t_2)
  \rangle = {\cal D}_p\,  \delta(t_1-t_2)\; ,
\end{equation}
the momentum diffusion coefficient can be identified with the
coefficient of the function $\delta(t_1-t_2)$.  Note that here we use
a different approach to the diffusion coefficient as compared to
earlier work \cite{horak97,hechenblaikner98,cohen92,domokos01}. It has
the additional advantage that the motional diffusion is linked
directly to the quantum noise accompanying the dissipation processes
of the system, i.e., the dipole fluctuations due to spontaneous
emission ($\xi_A$ that scales with $\gamma$) and the cavity photon
loss ($\xi_i$ that scales with $\kappa$). It allows to reveal how the
basic noise sources of the internal dynamics yield diffusion of the
external CM motion.

Finally, for the friction coefficient $\beta$, the dependence of the
internal variables $a_i$ and $\sigma$ on the velocity has to be taken
into account. Indeed, the non-conservative friction force arises from
the time-delayed reaction of the internal variables to the
displacement of the atom. They require a period of $\kappa^{-1}$ or
$\gamma^{-1}$ (in the coupled system typically the shorter time,
though this depends also on the coupling strength) to reach the
stationary state adapted to the actual position of the atom.  If the
atom moves much less than the wavelength during this time, i.e.~in the
low-velocity limit $k v \ll \gamma, \kappa$, the velocity-dependence
of the variables can be taken into account in a consistent way. The
time derivative is to be replaced by $d/dt \rightarrow
\partial/\partial t + v \partial/\partial x$ and, simultaneously, the
variables are expanded into the series
\begin{align}
\sigma &= \sigma^{(0)} + v \sigma^{(1)} + {\cal O}(v^2) \nonumber\\
a_i &= a_i^{(0)} + v a_i^{(1)} + {\cal O}(v^2) \nonumber\; .
\end{align}
The resulting dynamical equations can be solved systematically in
different orders of the velocity $v$. The first order components obey
the set of linear equations
\begin{subequations}
 \label{eq:HL1}
\begin{align}
 &(i \Delta_A - \gamma) \sigma^{(1)} - \sum_{i=1}^M g_i(z) a_i^{(1)}
 = \partial_z \sigma^{(0)} \\
 &(i\Delta_C - \kappa) a_i^{(1)} + g_i^*(z) \sigma^{(1)} = 
  \partial_z a_i^{(0)} \; ,
\end{align}
\end{subequations}
The coefficient $\beta$ of the linear friction force can be defined as
\begin{equation}
  \label{eq:beta}
  \beta/m=\langle F^{(1)} \rangle = i\hbar \sum_{i=1}^M 
     (\partial_z g_i) \langle {\sigma^\dagger}^{(0)} a_i^{(1)} + 
     {\sigma^\dagger}^{(1)} a_i^{(0)}\rangle - \mbox{H.c.} \; .
\end{equation}

\section{Solving the linear system}

In this section we perform the calculation of the linear friction and
the diffusion coefficients by directly solving the equations describing the
internal dynamics, i.e.\ Eqs.\ (\ref{eq:HL}) and (\ref{eq:HL1}).
Let us introduce the Fourier transform of the operators, 
\begin{equation}
  \label{eq:Fourier}
  O(t)=\frac{1}{\sqrt{2\pi}} \int_{-\infty}^{\infty} 
d\Omega e^{-i\Omega t} O(\Omega)
\end{equation}
and use the rule $\partial/\partial t \rightarrow -i\Omega$. The
velocity-independent part of the system variables yields
\begin{widetext}
\begin{subequations}
\label{eq:solution}
\begin{align}
\sigma^{(0)}(\Omega) &= \frac{1}{D(\Omega)} \left[(\kappa-i\Delta_C-i\Omega)
(\sqrt{2\pi} \delta(\Omega) \eta + \xi_A(\Omega) ) - \sum_{k=1}^{N}
g_k(z) \xi_k(\Omega) \right] \; ,\\
 a_i^{(0)}(\Omega) &= \frac{1}{D(\Omega)} \left[g_i^*(z) (\sqrt{2\pi}
 \delta(\Omega) \eta + \xi_A(\Omega) ) +
 \frac{D(\Omega)}{\kappa-i\Delta_C-i\Omega} \xi_i(\Omega) - \sum_{k=1}^{N}
\frac{g_i^*(z) g_k(z)}{\kappa-i\Delta_C-i\Omega} \xi_k(\Omega) \right] \; ,
\end{align}
\end{subequations}
\end{widetext}
where the determinant of the linear system reads
\begin{equation}
  \label{eq:det}
  D(\Omega)=(\kappa-i\Delta_C-i\Omega)(\gamma-i\Delta_A-i\Omega) +
  \sum_{k=1}^{N} |g_k(z)|^2 \; .
\end{equation}
The system variables are obtained as linear combinations of the
pumping term $\eta$ and noise operators. Note that, as a consequence
of the coupled dynamics, both the field-amplitude operators $a_i$ and
the dipole operator $\sigma$ incorporate the fluctuations $\xi_A$ and
$\xi_i$, associated with the spontaneous emission and from the cavity
photon loss, respectively.

In any normal-ordered expression of the operators $a_i^\dagger$,
$\sigma^\dagger$, $\sigma$, and $a$, all the noise operators $\xi$ are
found on the right side, while all the adjoint operators
$\xi^\dagger$ occur on the left side of the expression.
Therefore, when evaluating the mean value of a normal-ordered
product, the noise terms have no contribution.  It is useful to
introduce the c-number variables which arise from the coherent driving
terms of the exact expression (\ref{eq:solution}). They correspond to
the stationary solution of a semiclassical model of the same system,
and read
\begin{subequations}
  \label{eq:semiclass}
\begin{align}
  s^{(0)} &= \eta \frac{\kappa-i\Delta_C}{D} \; ,\\
  \alpha_i^{(0)} &= \eta \frac{g_i^*(z)}{D} \; ,
\end{align}
\end{subequations}
where, and hereafter in the paper, $D=D(0)$.  In normal-ordered
products, the operators can be replaced by these simple
``semiclassical'' solutions. For example, the atomic excitation is
obtained as
\begin{equation}
  \label{eq:popinv}
  \langle {\sigma^\dagger}^{(0)} \sigma^{(0)} \rangle = {s^{(0)}}^*
  s^{(0)} = \eta^2 (\kappa^2+\Delta_C^2)/|D|^2 \; ,
\end{equation}
which has to be well below one in order to be consistent with the
low-saturation assumption. The steady-state photon number in the mode $i$
is
\begin{align}
  \label{eq:photon}
  \langle {a_i^\dagger}^{(0)} a_i^{(0)} \rangle &= {\alpha^{(0)}}^*
  \alpha^{(0)} = \eta^2 |g_i(z)|^2/|D|^2 \nonumber\\
   &=\langle {\sigma^\dagger}^{(0)} \sigma^{(0)} \rangle \;
   |g_i(z)|^2/(\kappa^2+\Delta_C^2)\; .
\end{align}
The second expression exhibits how the photon number is related to the
atomic excitation. As this latter is necessarily small, the photon
number has to be below $g/\kappa$.

\subsection{The friction coefficient}

The friction coefficient is defined, in Eq.~(\ref{eq:beta}), by a
normally ordered expression.  Hence, it is enough to take into account
the coherent driving terms, i.e., the semiclassical solution
(\ref{eq:semiclass}) of the variables. Accordingly, the components,
first-order in velocity, have to be determined from the
Eqs.~(\ref{eq:HL1}) by using only the c-number part of the solutions.
This simplifying fact is in accordance with the physical intuition.
The non-adiabatic dynamics of the internal variables, being the origin
of the damping, does not substantially depend on the noise. On the
other hand, the non-adiabaticity must be reflected in the dynamics of
the c-number semiclassical variables, since it includes the pumping
and damping processes of the coupled internal system.

The first-order components of the variables can be easily
obtained. They are
\begin{subequations}
\begin{align}
    s^{(1)} &= \frac{\eta}{D^3} [(\kappa-i\Delta_C)^2-G] (\partial_z G)
     + \frac{\eta}{D^2} \Gamma \\
    \alpha_i^{(1)} &=  \frac{\eta}{D^3}
    (\kappa-i\Delta_C+\gamma-i\Delta_A) g_i^* (\partial_z G) \nonumber\\
    & \qquad - \frac{\eta}{D^2} \frac{1}{\kappa-i\Delta_C} \left(D (\partial_z g_i^*)
     -g_i^* \Gamma \right) \; ,
\end{align}
\end{subequations}
where 
\begin{subequations}
  \label{eq:G}
  \begin{align}
    G &= \sum_{k=1}^M |g_k|^2 \\
    \Gamma &= \sum_{k=1}^M g_k (\partial_z g_k^*)\; ,
  \end{align}
\end{subequations}
Note that $(\partial_z G)=\Gamma+\Gamma^*$. Replacing these
expressions into the definition (\ref{eq:beta}), one gets the friction
coefficient. Here we present only the solution for a standing-wave
cavity, where the mode functions are real. It is
\begin{widetext}
\begin{equation}
  \begin{split}
    \label{eq:friction_gen}
    \beta = - &\hbar \frac{\eta^2}{|D|^4} (\partial_z G)^2 \left[\kappa \Delta_A
      + 2 \Delta_C
      \left(\kappa+\gamma+\kappa \frac{\kappa\gamma -\Delta_C\Delta_A+G/2}{\kappa^2+\Delta_C^2} \right)\right] \\
    + &4 \hbar \frac{\eta^2}{|D|^6} (\partial_z G)^2 \Delta_C
    (\kappa\Delta_A+\gamma\Delta_C) [\kappa^2 \Delta_A+
    \gamma^2 \Delta_C +
    (\Delta_A+\Delta_C) (\Delta_A\Delta_C-G)] \\
    + & 4 \hbar \frac{\eta^2}{|D|^2}
    \frac{\kappa\Delta_C}{\kappa^2+\Delta_C^2} \sum_k (\partial_z g_k)^2
  \end{split}
\end{equation}
\end{widetext}
The generalization for running-wave modes in a ring cavity is
straightforward.

\subsection{Diffusion coefficient}

The diffusion stems from the fluctuation of the force operator, i.e.,
that of the dipole interaction term of the Hamiltonian (\ref{eq:H}).
When calculating the diffusion coefficient from its definition
(\ref{eq:diffusion_def}), the product of the force operators contains
non-normally ordered terms. Hence, the noise terms in the solutions
(\ref{eq:solution}) have non-vanishing contributions, and the noise
correlations given in (\ref{eq:L}) have to be used in the calculation.
One can separate terms originating from the spontaneous emission noise
($\xi_A$) and from the cavity loss noise ($\xi_i$). We omit the
details of the lengthy calculation here apart frommentioning one 
non-trivial step in the derivation. Although the sources $\xi_A$
and $\xi_i$ ($i=1...M$) are supposed to represent white noise, the
resulting noise associated with the force operator becomes coloured,
that is, it has a non-uniform spectrum. Accordingly, in addition to
the Dirac-$\delta$ correlation assumed in the definition
(\ref{eq:diffusion_def}), one obtains other terms proportional to the
derivatives of the Dirac-$\delta$ to all order. We neglect these terms
and consider the coefficient of the Dirac-$\delta$ to describe the
diffusion process.

The diffusion coefficient for real mode functions reads
\begin{multline}
    \label{eq:diff_gen}
      {\cal D}_{\mbox{\tiny dip}} = 2 \hbar^2 \frac{\eta^2}{|D|^2}
 \Biggl(\kappa \sum_k \left( \partial_z g_k \right)^2 \\
+ (\partial_z G)^2 \Delta_C \frac{\kappa\Delta_A+\gamma\Delta_C}{|D|^2} \Biggr)
\end{multline}
In addition to the dipole heating, the noise induced by the random recoil
accompanying the spontaneous
emission has to be taken into account. The recoil contributes to the total
diffusion by
\begin{equation}
  \label{eq:diff_recoil}
  {\cal D}_{\mbox{\tiny rec}} = 2 \hbar^2 k_A^2 \bar{u^2} \frac{\eta^2}{|D|^2} (\kappa^2+\Delta_C^2)
  \gamma\; ,
\end{equation}
where $k_A=\omega_A/c$ and $\bar{u^2}$, characteristic of the atomic
transition, is the mean of the recoil momentum projected on the cavity
axis.

To conclude this section let us emphasize that the results for the
friction and the diffusion coefficients have a general validity
regardless of the relationship of the parameters involved. The only
condition is the weak atomic excitation that can always be met by a
proper adjustment of the pumping strength $\eta$.

\section{Cooling regimes}

In this section we evaluate the previously calculated expressions for
the diffusion and friction coefficients for a single-mode cavity.
Such an analysis serves as a ground to identify the physical processes
underlying the cavity-cooling. In addition, it is interesting to
compare the results with those obtained in a similar system
\cite{horak97,hechenblaikner98} with the cavity mode being driven
instead of the atom.

\begin{figure}[htbp]
  \begin{center}
    \includegraphics[width=5cm]{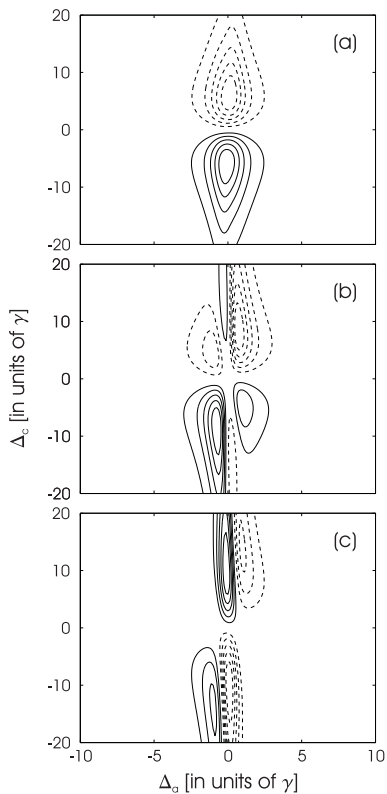}
    \caption{Contour plot of the friction coefficient in the
      bad-cavity regime ($\kappa=10\gamma$) for various coupling
      constants, (a) $g=\gamma/2$, (b) $g=3\gamma/2$, (c) $g=3\gamma$,
      linking the closely independent atom and cavity field system to
      the dressed atom one. Solid contour lines indicate cooling
      ($\beta<0$), dashed ones heating ($\beta>0$) regions.}
    \label{fig:fric_gvar}
  \end{center}
\end{figure}

\begin{figure}[htbp]
  \begin{center}
    \includegraphics[width=5cm]{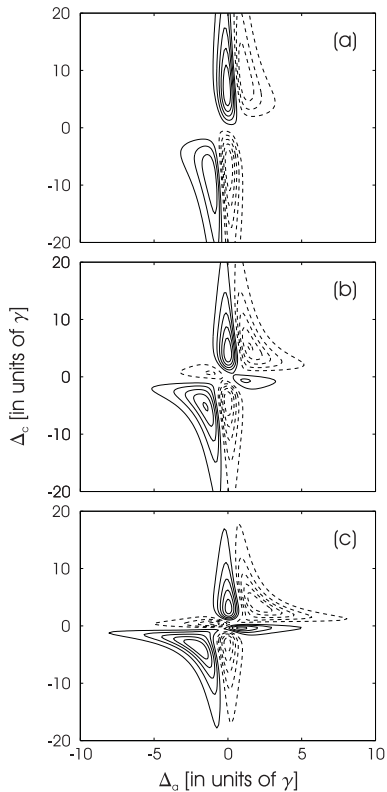}
    \caption{Contour plot of the friction coefficient in the
      dressed-atom regime ($g=3\gamma$) for decreasing cavity
      linewidth, (a) $\kappa=5\gamma$, (b) $\kappa=2\gamma$, (c)
      $\kappa=\gamma$, linking the adiabatic field dynamics regime to
      the one including time-delay, and the accompanying
      non-conservative cavity forces.}
    \label{fig:fric_kapvar}
  \end{center}
\end{figure}

There were several attempts to interpret the cooling in terms of
simple physical processes, such as two-photon Doppler effect
\cite{vuletic99}, or Sisyphus-effect in the dressed state basis
\cite{hechenblaikner98}. In what follows, we propose to explore the
friction mechanism by systematically varying the parameters, through
relatively simple limiting cases to a final, quite general parameter
setting in the strong-coupling cavity QED regime. In Figs.\ 
\ref{fig:fric_gvar} and \ref{fig:fric_kapvar} the friction coefficient
is shown as a function of the detunings in a contour plot style for
various characteristic values $\kappa$ and $g$ of the cavity.

Let us start in the bad-cavity limit with a negligibly small coupling
parameter ($\kappa= 10\gamma$, $g=\gamma/2$). This corresponds to the
simple perturbative regime of cavity QED, where the role of the cavity
is that it reshapes the radiative environment of the atom and
increases the spontaneous emission rate at the cavity frequency.  In
our one-dimensional model we are interested in mechanical forces due
to cavity photons. As these photons leak out of the cavity fast, such
that no dynamics occur on a time scale longer than $\kappa^{-1}$, the
atom-cavity field interaction can be basically considered as
scattering. Although the atomic dipole is linear in the low-saturation
regime, inelastic scattering can happen due to the CM motion that is
able to compensate for the energy difference. For $\omega_C>\omega_p$,
the spontaneous emission being favoured at the cavity frequency higher
than the incoming one, the scattering is accompanied by a loss of the
kinetic energy, i.e.\ cooling.  This is the origin of the cooling
region for $\Delta_C<0$ in Fig.~\ref{fig:fric_gvar}a, and in turn, the
same mechanism acts reversly leading to heating for $\Delta_C>0$.
Note that this simple energy conservation argument is also in the
heart of the interpretation given in \cite{vuletic01} for the
two-photon Doppler effect.

Next, let us keep $\kappa$ large and increase gradually the coupling
constant $g$. Figures~\ref{fig:fric_gvar}b and c correspond to
$g=1.5\gamma$ and $g=3\gamma$, respectively. Two sharp peaks emerge at
$\Delta_A\approx 0$.  As $\kappa$ is still the far largest parameter,
the scattering picture applies.  However, instead of the bare atom,
the strongly coupled atom-cavity system has to be taken into account.
As a consequence, relatively close to the pumping frequency
$\omega_p$, the spectrum now exhibits the doublet of the first excited
dressed-state manifold. Scattering of a pump photon into a cavity one
may happen now via both intermediate states.  In the case of
$\Delta_C<0$ and $\Delta_A\approx 0$, for example, which is the lower
half plane of the plots \ref{fig:fric_gvar}b and c, the upper dressed
state $| + \rangle$ is at about the cavity frequency $\omega_C$ with a
width of $\kappa$. The upper level therefore amounts to cavity photons
of frequency about $\omega_C$, which corresponds to the process we
described just before and is responsible for the broad cooling peak in
the background.  For large enough coupling constant, however, the
lower state $|-\rangle$ is contaminated with a non-negligible amount
of $|g,1\rangle$ component, its weight is proportional to
$g^2/(\omega_C-\omega_A)^2$.  Pumping photons can then be transmitted
into the cavity by exciting the lower dressed state at the frequency
of about $\omega_A + 2 g^2/(\omega_A-\omega_C)$. For $\Delta_A\approx
0$ this state is more resonant with the pump and the corresponding
scattering channel becomes dominant. That is, one gets the heating
peak for $\Delta_C<0$ and $\Delta_A > 0$, whose width is approximately
$\gamma$, the one of the lower dressed level.  For $\Delta_A<0$ the
driving field is tuned below the lower dressed level and thus
scattering via this state yields cooling again. Note the displacement
of the heating peak with respect to the axis $\Delta_A=0$ which is due
to the cavity-vacuum induced lightshift, i.\ e., the lower
dressed-state resonance and the atomic resonance do not coincide.

Regardless of the value of the coupling constant, as long as $g\ll
\kappa$, the field adiabatically follows the atomic dynamics, as is
the case for the parameter settings chosen for
Fig.~\ref{fig:fric_gvar}. The damping effect cannot be attributed to a
non-conservative, time-delayed field dynamics.  This kind of behaviour
occurs when decreasing the cavity linewidth $\kappa$. Then, instead of
being a passive element with specific mode density, it becomes
necessary to include the cavity field as a dynamical component of the
system. For longer spontaneous lifetimes, the system spends some time
in the dressed states and the slow atom moves in a potential
associated with the sinusoidally varying dressed levels.

In Figs.~\ref{fig:fric_kapvar}a--c, the case of $\kappa=5\gamma$,
$2\gamma$, and $\gamma$ are shown, respectively. There appears a pair
of peaks with hyperbolic shape along $\Delta_C\Delta_A \approx g^2$,
which is cooling for negative detunings $\Delta_C, \Delta_A <0$, and
heating for positive ones $\Delta_C, \Delta_A >0$. This cooling
(heating) region corresponds to the situation where the lower (upper)
dressed state is resonantly pumped at antinodes (maximum coupling) and
thus the slowly moving atom has to, on average, climb up potential
hills (descend to potential valleys). 

Finally, there are two peaks at at $\Delta_C\approx 0$ that can be
attributed to the effect of the Doppler-shift in the correlated
atom-field dynamics. Compared to the free-space Doppler cooling case,
the preferential direction appears in the emission rather than in the
absorption. For $\Delta_A>0$ and $\Delta_C \approx 0$, the atom gets
closer to resonance with the copropagating component of the
standing-wave field, i.e., following an emission the atom is more
likely to impart a recoil in the direction opposite to its velocity.

One can expect that in the low-saturation regime, where the
Heisenberg-Langevin equations are linearized, it makes no difference
which component of the coupled system is being pumped. Indeed, this
intuition is reflected in the mathematical form of the results. We
checked that a systematic exchange of the parameters ($\Delta_A$,
$\gamma$) $\leftrightarrow$ ($\Delta_C$, $\kappa$) in the expressions
(\ref{eq:friction_gen}) and (\ref{eq:diff_gen}) reproduces the results
obtained for the cavity-driven case in Ref.\ \cite{hechenblaikner98}.
This symmetry reveals that the roles of the two oscillators, the field
mode and the atomic dipole, are interchanged.  Accordingly, the map of
the friction coefficient in Fig.\ \ref{fig:fric_kapvar}c is similar to
the Fig.\ 3 of Ref.\ \cite{hechenblaikner98} with the detuning axes
exchanged (reflection to the diagonal).  In principle, for
$\kappa=\gamma$ there is a one-to-one correspondence between the two
systems, i.e. a given dynamics of the atom-driving configuration can
be mimicked, with exchanged detunings $\Delta_C$ and $\Delta_A$, in
the cavity-driving one.  However, for a fixed setting of the
detunings, the accompanying cooling (or heating) mechanism is
different, which leads to an essential modification of the relevant
physical quantities. This deviation becomes of importance when there
is an additional constraint concerning the detunings, for example,
$\Delta_A$ has to be very large to keep the spontaneous photon
scattering low for molecule cooling.  The more detailed analysis of
this comparison is relegated into the next section, where also other
numerical examples for the characteristic statistical features are
presented.

\section{Temperature and cooling times in a multimode cavity}

In this section we will study thermodynamic properties of the system.
Most importantly, we calculate the steady-state temperature which can
be estimated by the ratio of the spatially averaged diffusion and
friction coefficients. Localization effects were proven
\cite{domokos01} to be important for cavity fields with higher
intensity than the sub-photonic fields occuring in the present
scenario. Hence, uniform position distribution of the atom can safely
be assumed. In this approximation, the temperature becomes independent
of the pumping strength $\eta$. The other important feature is the
time scale of reaching the given steady-state temperature. The
so-called cooling time can be considered to be the inverse of the
friction coefficient $\beta$.  However, it depends on the pumping
strength $\eta$ which is quite arbitrary, of course, within the limit
of not to excite the atomic dipole too much. The interesting, pumping
independent quantity, in fact, is the number of spontaneously
scattered photons during the cooling time $\beta^{-1}$. Low number of
spontaneous emission means efficient cooling, where the cooling time
scale due to the cavity dissipation channel is short enough compared
to the spontaneous scattering into lateral modes. In the following
analysis we include the scaling of these two quantities on the number
of degenerate modes of the cavity.

\subsection{Effective mode approach}

We will study this effect first in a simplified geometry when the
spatial variation of the different modes in the degenerate manifold is
closely uniform. This can happen, for example, with a piece of coated
waveguide where many quasi-degenerate sinusoidal modes can be found
within the atomic spectral linewidth with slightly different wave
numbers.  In the following, we will consider another example, the
first few transverse modes in a confocal cavity. The transverse modes
with even index are exactly degenerate.  The corresponding mode
functions, the Hermite-Gaussian modes, are known in the paraxial
approximation.  To be conform with it, the maximum transverse mode
indices we can take into account are limited by
$(n+m)\lambda/l_{\mbox{\tiny cav}} \ll 1$, where $l_{\mbox{\tiny
    cav}}$ is the cavity length. This limitation also implies that the
mode functions can be simplified around the cavity center $z=0$.
First, the Guoy phase term can be omitted, and the variation of the
longitudinal wave number $k_{2n,2m}=k_{0,0}-2(2n+2m+1)/l_{\mbox{\tiny
    cav}}$ can be neglected, i.~e.  $k_{2n,2m} \approx k_{0,0}\equiv
k$.  Accordingly, the mode function along the cavity axis can be
approximated by the simple cosine function $\cos{(k z)}$.  Second,
only the leading term of the derivative $\partial_z g(z)$, expanded
into a power series of $\lambda/l_{\mbox{\tiny cav}}$, must be kept,
which is proportional to $k \sin{(kz)}$. In a region close to the
cavity axis the mode functions thus overlap and form an effective mode
with enhanced coupling to the atom. It is an interesting problem to
move out from this limit into a situation where the different modes
have highly varying derivatives in space, that one has to study in a
three dimensional context with the exact mode functions.

In the simple example described above the presence of many degenerate
modes can be incorporated in an effective coupling constant
$g_{\mbox{\tiny eff}}$, a concept already used in Ref.\ 
\cite{fischer2001}.  The enhancement factor is
\begin{equation}
  \label{eq:g_eff}
  g_{\mbox{\tiny eff}}/g = \frac{(2N+1)!!}{(2N)!!}\; ,  
\end{equation}
where $2N$ is the maximum index taken into account, that is we
consider a total number of modes $M=(N+1)^2$ in the dynamics. The
effective coupling constant $g_{\mbox{\tiny eff}}$ grows closely as a
linear function of $N$, which indicates that orders of magnitude can
be gained in the coupling strength. The unphysical divergence for
large $N$ stems from the extension of the mode functions obtained
within the paraxial approximation to high indices of $N$. In practice,
the effective $g$ could be measured and then an effective number of
modes can be determined.

\begin{figure}[ht]
  \begin{center}
    \includegraphics[width=7cm]{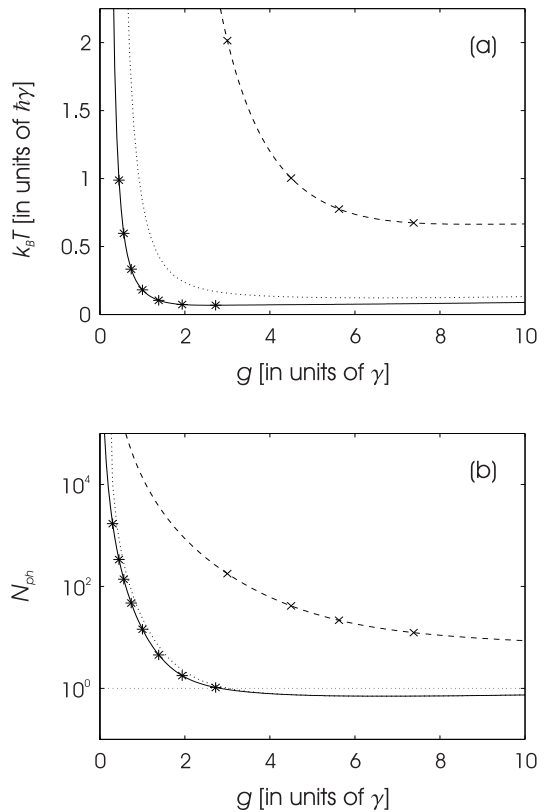}
\caption{Steady-state temperature (a), and the number of
  spontaneously scattered photons in a cooling time (b) versus the
  effective coupling constant $g_{\mbox{\tiny eff}}$. Dashed line
  corresponds to $\kappa=\gamma$, the solid one to $\kappa=\gamma/10$.
  The crosses put on the curves indicate the coupling constant
  corresponding to maximum transverse indices of $2N = 0, 2, 4, 8$, ..., with
  starting a single-mode coupling constant of $g=3\gamma$ for the
  dashed curve of $\kappa=\gamma$, and $g=3\gamma/10$ for the solid
  curve of $\kappa=\gamma/10$. For this latter the single-mode
  coupling (maximum index 0) is missing from the plotted range of
  temperatures.}
    \label{fig:multi_reg2}
  \end{center}
\end{figure}

Let us now see how the steady-state properties of the system depend on
the number of modes in the confocal cavity example, i.e., on the
effective coupling constant. The steady-state temperature is plotted
in Fig.\ \ref{fig:multi_reg2}a for the hyperbolic cooling regime with
$\Delta_C, \Delta_A < 0$, and $\Delta_C\approx 0$. When varying
$g_{\mbox{\tiny eff}}$, the detunings $\Delta_C$ and $\Delta_A$ are
rescaled such that their product is fixed at $g_{\mbox{\tiny eff}}^2$,
and their difference is constant. The first of these conditions
ensures that the lower dressed state is pumped resonantly at an
antinode (minimum energy). The latter one means that only the pumping
frequency $\omega_p$ is to be varied, both the atomic $\omega_A$ and
the degenerate mode frequency $\omega_C$ are fixed. For the plot we
set $\Delta_A-\Delta_C = \omega_C-\omega_A = - 50 \gamma$. Since
$\Delta_A \approx - 50 \gamma$, i.e., the driving field is tuned to be
very far from resonance.  The cavity properties are partly determined
by the parameter $\kappa$. The solid curve in the figure corresponds
to $\kappa=\gamma$, the dashed one to $\kappa=\gamma/10$. The other
relevant cavity parameter $g_{\mbox{\tiny eff}}$ is considered a
variable, however, it is useful to define the single-mode coupling
constant $g$. It is then set to $g=3\gamma$ and $g=3 \gamma /10$,
respectively, as if the cavity length were changed by a factor of 10,
keeping the same mirror transmittivity. Having defined the single-mode
coupling $g$, a discrete series of effective coupling constants is
obtained for increasing number of modes, which is indicated by the
points on the curves.
The rapid drop in temperature obtained when the number of modes
included slightly increases is the main benefit we can expect from the
use of a multimode cavity. In both cases, $\kappa=\gamma$ and
$\kappa=\gamma/10$ sub-Doppler temperatures can be achieved with a
small number of modes involved. Note also that the curves converge
fast, indicating the existence of a well-defined temperature
insensitive to the divergence of the effective coupling constant
$g_{\mbox{\tiny eff}}$ with increasing number of modes.

In addition to the induced atomic dipole moment, the driving field
yields a small, but not completely negligible atomic excitation.
Hence, spontaneous photon scattering occurs with a rate of $2 \gamma
\langle \sigma^\dagger \sigma \rangle$. It is an important quantity
how many photons are scattered in this way during the characteristic
time of cooling which is $(2 \beta)^{-1}$ in our case. The principal
goal is to restrict this number below one which implies that the
scheme can be extended for cooling particles without closed pumping
cycle. The Fig.\ \ref{fig:multi_reg2}b shows that the number of
spontaneously scattered photons in a cooling time, $N_{\mbox{\tiny
    ph}}= \gamma \langle \sigma^\dagger \sigma \rangle/\beta$, can
decrease below the limit of one photon for large enough
$g_{\mbox{\tiny eff}}$, corresponding to about $N=64$.

Both quantities plotted in Fig.\ \ref{fig:multi_reg2} are
independent of the pumping strength. The cooling time itself depends
on it. However, without specifing a driving field intensity, one can
deduce numerical values of the cooling time from the Fig.\ 
\ref{fig:multi_reg2}b, provided the saturation is kept fixed. For
example, at a saturation $\langle \sigma^\dagger \sigma \rangle\approx
0.1$, and for Rb with $\gamma=20/\mu$s, the cooling time is
$N_{\mbox{\tiny ph}}/4$ in units of $\mu$s.

The result exhibited in Fig.\ \ref{fig:multi_reg2} suggests that
smaller $\kappa$ provides better performances in terms of temperature,
cooling time. On the other hand, as it was pointed out in
\cite{vuletic01,horak01}, the velocity capture range is limited by $k
v < \kappa$.

Finally, let us return to the problem already addressed in the last
section. Fig.\ \ref{fig:multi_reg2} presents an additional curve
(dotted line) that corresponds to the same parameter setting as the
one belonging to the solid line ($\kappa=\gamma/10$), however, the
single-mode cavity field is being pumped instead of the atom. Whatever
component is driven externally, the pumping field, by construction of
the detunings, is resonant with the lower dressed state at an
antinode, and this analogy makes the comparison justified. It is
apparent that the temperature in the atom-driven case is lower. The
curve belonging to the cavity-driven case for $\kappa=\gamma$ does not
even fit in the plotted range, the difference with respect to the
dashed line is much larger. Although a simple transformation connects
the results of the atom- and cavity-driven cases, as this example
illustrates it, a significant physical difference can occur. The
origin is that the detunings were the same, $\Delta_A$ is large and
$\Delta_C$ is small, which breaks the symmetry between the two systems
based on the interchange of the above detunings.

\subsection{Beyond the effective mode approach}

The effective mode approach could be used in the previous analysis
because all the relevant modes closely overlapped in the region of
interest, i.~e., around the cavity center. Accordingly, the system is
reminiscent of a single-mode one with enlarged coupling constant
$g_{\mbox{\tiny eff}}$. By contrast, when one moves away from the
center, but still on the axis, the cosine-like mode functions with
varying wavenumbers undergo a dephasing. One immediate consequence is
that the friction force, proportional with the gradient of the mode
function, does not vanish in any point. Furthermore, on inspecting the
general solutions (\ref{eq:friction_gen}) and (\ref{eq:diff_gen}), one
can notice that the friction coefficient $\beta$ is proportional to
the square of the sum $\partial_z \sum g_k^2(z)$, while the diffusion
coefficient, for $\Delta_C \cong 0$ contains only the sum $\sum (\partial_z
g_k)^2$.  Although the determinant $|D|^2$ appears also on different
powers in the denominator, it is clear that for certain parameter
settings the two coefficients can scale in a different way with the
number of modes.  This gives rise to the possibility to get an
interferometric enhancement of the cooling by a collective effect of
the modes and leads us to conjecture that the steady-state temperature
can be lower in other positions than the cavity center.

\begin{figure}[ht]
  \begin{center}
    \includegraphics[width=7cm]{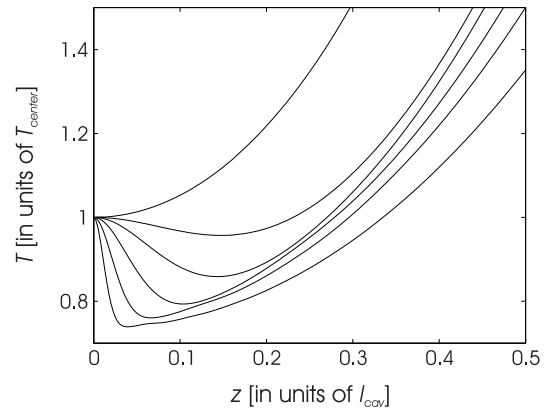}
\caption{Steady-state temperature as a function of the position in the
  cavity for $\kappa=\gamma/10$, $g=3\gamma/10$. The temperature in
  the center is used as a unit, therefore purely the competitive
  effects of the dephasing and the decrease of the coupling strength
  when moving away from the waist can be observed. The curves correspond
  0, 2, 4, 8, 16 and 32 as maximum index taken into account.}
    \label{fig:range}
  \end{center}
\end{figure}

To check this expectation, we calculated the temperature as a function
of the position in the cavity by using the mode functions
$\cos\{(kz-(2m+2n+1)\mbox{atan}(z/z_0)\}$ with transverse indices
$n,m=0, 1, N$, where the Guoy-phase term is responsible for shifting
the modes out of phase. The length scale of the dephasing can be
estimated by $z/l_{\mbox{\tiny cav}} \sim \pi/4(2N+1)$, i.\ e., at
this distance from the center the Guoy-phase shift is about $\pi/2$.
For each position, (i) we perform again the averaging of the diffusion
and friction coefficients over a wavelength, and (ii) we redefine the
detunings $\Delta_A$, $\Delta_C$ such that the lower-dressed state
corresponding to the local coupling constant be resonantly pumped at
antinodes. The result is plotted in figure \ref{fig:range}.
The temperature initially drops as the atom moves out from the center.
The higher indices are taken into account, the faster the initial drop
happens, which suggests that the underlying reason is indeed the
dephasing of the cosine modes. The estimated length scale shows a good
agreement with the numerically obtained results for various maximum
indices $2N$. The reduction of about $20 \%$ in the temperature can be
attributed to a collective, interference-like effect of the multimode
field. For large distances from the center, after the dephasing is
completed, the temperature grows slowly again, exhibiting the effect
of the decreasing coupling constant.

\section{Conclusions}

The mechanical effects of a high-$Q$ cavity on the center--of--mass
motion of a coherently-driven neutral atom have been investigated. We
calculated the diffusion coefficient and the friction force from a
quantum model adapted to the strong atom-field coupling regime, hence
the validity does not depend on any specific relationship of the
parameters. The model is analogous to the one described in Refs.\ 
\cite{horak97,hechenblaikner98}, however, we considered the system
with the atom being externally pumped instead of the cavity field
mode. Surprisingly, this difference leads to important new features in
the diffusion and damping process.  Lower temperatures can be achieved
in the present system. We pointed out that certain cooling mechanisms
can be realized only in the good-cavity limit, i.e.\ $g \gg \kappa,
\gamma$, where the dressed-atom dynamics including Rabi oscillations
becomes dominant. The corresponding parameter regimes are especially
suited to applications, since large atomic detunings (red or blue) are
allowed here. Furthermore, we found that drastic improvement in terms
of low temperature and small number of spontaneous scattering can be
obtanied in a degenerate multimode resonator. As a highlight of this
possible benefit, we showed that the number of spontaneously emitted
photons from the atom during the cooling time can be reduced to below
one, which demonstrates the possibility of cooling molecules optically
below the Doppler limit.

\acknowledgments

We thank Peter Horak, Pepijn Pinkse, Gerhard Rempe, and Vladan
Vuleti\'c for fruitful discussions.  This work was supported by the
Austrian Science Foundation FWF (Project P13435).  P.~D.\ acknowledges
the financial support by the National Scientific Fund of Hungary under
contracts No. T034484 and F032341.

\end{document}